\RequirePackage{fix-cm}
\documentclass[twocolumn,epjc3,comma,sort&compress,natbib]{svjour3}
\bibpunct{[}{]}{,}{n}{}{,} 

\journalname{Eur. Phys. J. A}

\usepackage{latexsym}
\usepackage{amsmath}
\usepackage{amssymb}
\usepackage{amsfonts}

\usepackage[mathscr,scaled=1.15]{urwchancal}
\DeclareFontFamily{OT1}{pzc}{}
\DeclareFontShape{OT1}{pzc}{m}{it}%
{<-> s * [1.15] pzcmi7t}{}
\DeclareMathAlphabet{\mathpzc}{OT1}{pzc}{m}{it}

\usepackage{color}

\usepackage{supertabular}
\usepackage{placeins}
\usepackage{epsfig}
\usepackage{graphicx}

\definecolor{purple}{rgb}{0.5,0,0.5}
\definecolor{blue}{rgb}{0.0,0,0.9}
\definecolor{prdblue}{rgb}{0.133,0.118,0.498}
\usepackage[colorlinks=true, pdfstartview=FitV, linkcolor=prdblue, citecolor= prdblue, urlcolor=prdblue]{hyperref}

\begin{document}

\title{$\,$\\[-7ex]\hspace*{\fill}{\normalsize{\sf\emph{Preprint no}. NJU-INP 019/20}}\\[1ex]
Higgs modulation of emergent mass as revealed in kaon and pion parton distributions
}

\author{Z.-F. Cui\thanksref{eZFC,NJU,INP}
        \and
       M. Ding\thanksref{eMD,ECT} 
       \and
       F. Gao\thanksref{eFG,UH}
       \and
       K. Raya\thanksref{eKR,NKU}
       \and
       D. Binosi\thanksref{eDB,ECT}
       \and
       L. Chang\thanksref{eLC,NKU}
       \and
       C. D. Roberts\thanksref{eCDR,NJU,INP}
       \and
       J. Rodr\'{\i}guez-Quintero\thanksref{ePepe,UHE}
       \and
       S. M. Schmidt\thanksref{eSMS,HZDR,RWTH}
}

\thankstext{eZFC}{phycui@nju.edu.cn}
\thankstext{eMD}{mding@ectstar.eu}
\thankstext{eFG}{f.gao@thphys.uni-heidelberg.de}
\thankstext{eKR}{khepani@nankai.edu.cn}
\thankstext{eDB}{binosi@ectstar.eu}
\thankstext{eLC}{leichang@nankai.edu.cn}
\thankstext{eCDR}{cdroberts@nju.edu.cn}
\thankstext{ePepe}{jose.rodriguez@dfaie.uhu.es}
\thankstext{eSMS}{s.schmidt@hzdr.de}

\authorrunning{Z.-F. Cui \emph{et al}.} 

\institute{School of Physics, Nanjing University, Nanjing, Jiangsu 210093, China \label{NJU}
           \and
           Institute for Nonperturbative Physics, Nanjing University, Nanjing, Jiangsu 210093, China \label{INP}
           \and
           European Centre for Theoretical Studies in Nuclear Physics
            and Related Areas (ECT$^\ast$) and Fondazione Bruno Kessler\\
            \hspace*{0.5em}Villa Tambosi, Strada delle Tabarelle 286, I-38123 Villazzano (TN), Italy\label{ECT}
            \and
            Institut f{\"u}r Theoretische Physik, Universit{\"a}t Heidelberg, Philosophenweg 16, D-69120 Heidelberg, Germany\label{UH}
            \and
            School of Physics, Nankai University, Tianjin 300071, China\label{NKU}
            \and
            Department of Integrated Sciences and Center for Advanced Studies in Physics, Mathematics and Computation, \\ \hspace*{0.5em}University of Huelva, E-21071 Huelva, Spain\label{UHE}
            \and
            Helmholtz-Zentrum Dresden-Rossendorf, Dresden D-01314, Germany\label{HZDR}
            \and
            RWTH Aachen University, III. Physikalisches Institut B, Aachen D-52074, Germany\label{RWTH}
            }

\date{2020 November 18}

\maketitle

\begin{abstract}
Strangeness was discovered roughly seventy years ago, lodged in a particle now known as the kaon, $K$.  Kindred to the pion, $\pi$; both states are massless in the absence of Higgs-boson couplings.  Kaons and pions are Nature's most fundamental Nambu-Gold\-stone modes.  Their properties are largely determined by the mechanisms responsible for emergent mass in the standard model, but modulations applied by the Higgs are crucial to Universe evolution.  Despite their importance, little is known empirically about $K$ and $\pi$ structure.  This study delivers the first parameter-free predictions for all $K$ distribution functions (DFs) and comparisons with the analogous $\pi$ distributions, \emph{i.e}.\ the one-dimensional maps that reveal how the light-front momentum of these states is shared amongst the gluons and quarks from which they are formed.
The results should stimulate improved analyses of existing data and motivate new experiments sensitive to all $K$  and $\pi$ DFs.
\end{abstract}



\noindent\textbf{1.$\;$Introduction}.\,---\,%
%
The kaon was discovered in 1947 \cite{Rochester:1947mi}; yet, today, seventy years later, almost nothing is known about kaon structure.  (Regarding the pion, Nature's closest approximation to a Nambu-Goldstone (NG) mode \cite{Nambu:1960tm, Goldstone:1961eq}, the position is marginally better \cite{Horn:2016rip, Aguilar:2019teb, Roberts:2020udq}.)  This is unsatisfactory for many reasons.  Primary amongst them is the fact that the standard model of particle physics (SM) has two sources of mass: explicit, generated by couplings to the Higgs-boson; and emergent, a dynamical consequence of strong interactions, responsible for the $m_N \sim 1\,$GeV mass-scale that characterises nuclei and the origin of more than 98\% of visible mass.  Emergent hadronic mass (EHM) is dominant for all nuclear physics systems; but Higgs-induced modulations are critical to the evolution of the Universe, \emph{e.g}.\ CP-violation, discovered in neutral kaon decays \cite{Christenson:1964fg}.

Knowledge of kaon structure opens a window onto the interference between Higgs boson effects and EHM \cite{Aguilar:2019teb, Roberts:2020udq}, \emph{e.g}.\ within quantum chromodynamics (QCD), $\pi$ and $K$ mesons are identical without a Higgs mechanism: they are NG modes, whose common properties are determined by EHM.  When the Higgs coupling is switched on, the Lagrangian mass of the $s$-quark becomes $\approx 27$-times greater than the mean $u$, $d$ quark mass; yet, the ratio of $K$ and $\pi$ decay constants changes by only 20\%.  Herein, therefore,
we deliver parameter-free predictions for all kaon distribution functions: valence, glue and sea.  Our results are founded on two basic features of QCD:
%
concerning integrated quantities, reliable approximations may be obtained using a factorised form for a hadron's light-front wave function (LFWF) \cite{Xu:2018eii};
and
the existence of a process-independent effective charge provides both an unambiguous definition of the infrared scale that characterises nonperturbative calculations of hadron parton distributions and the basis for QCD evolution to the higher scales accessible to experiment \cite{Cui:2019dwv}.


\smallskip

\noindent\textbf{2.$\;$Light-front wave functions and parton distributions}.\,---\,%
%
Given a Poincar\'e-covariant Bethe-Salpeter wave function \cite{Nakanishi:1969ph}, $\chi_M(\bar k;P;\zeta_H)$, for a meson $M=\pi_{u\bar d}, K_{u\bar s}$, where $k_f$ is the momentum of the valence $f$-quark, $P=k_u-k_{\bar h}$, $2\bar k = k_u+k_{\bar h}$, the two-particle distribution amplitude (DA) for the $u$-quark can be obtained by light-front projection \cite{Chang:2013pq}:
\begin{subequations}
\begin{align}
&f_M \,\varphi^u_{M}(x;\zeta_H)= \frac{1}{16\pi^3}\int d^2 k_\perp \, \psi_{M_u}^{\uparrow\downarrow}(x,k_\perp^2;\zeta_H)
\label{LFWFexplicit}\\
& =N_c {\rm tr}Z_2(\zeta_H,\Lambda) \int_{dk}^\Lambda \delta_n^x(k_u)\gamma_5 \gamma\cdot n \chi_M(\bar k;P ;\zeta_H)\,, \label{varphiresult}
\end{align}
\end{subequations}
where $\psi_{M_u}^{\uparrow\downarrow}(x,k_\perp^2;\zeta_H)$ is the meson's two-particle LFWF;
$N_c=3$; the trace is over spinor indices; $\int_{dk}^\Lambda$ is a symme\-try-preserving regularisation of the four-dimen\-sional integral, with $\Lambda$ the regularisation scale; $Z_2$ is the quark wave function renormalisation constant; $\delta_n^x(k_u) = \delta(n\cdot k_u - x n\cdot P)$, $n^2=0$, $n\cdot P = -m_M$  in the meson rest frame, with $m_M$ the meson's mass; and $f_M$ is the meson's leptonic decay constant.  The DA for the $h$-antiquark is
$\varphi_{M}^{\bar h}(x;\zeta_H) =  \varphi_{M}^{u}(1-x;\zeta_H)$.
Here, the argument $x$ is the light-front fraction of the meson's momentum carried by the $u$-quark and $\vec{k_\perp}$ is the momentum of this quark in the associated transverse plane.

To solve the QCD bound-state problem, we choose a momentum subtraction renormalisation procedure with renormalisation scale $\zeta=\zeta_H$, \emph{i.e}.\ all quantities are renormalised at the infrared ``hadronic'' scale whereat the dressed quasiparticles obtained as solutions of the quark gap equation express all properties of the bound state under consideration \cite{Roberts:2015lja}, \emph{e.g}.\ they carry all the hadron's momentum at $\zeta_H$.
Stated differently, at $\zeta_H$, all bare-gluon and -quark contributions that would be explicit in a treatment of structure functions using a parton-basis Fock-space expansion are resummed, using nonperturbative quantum field equations of motion, into the compound quasiparticle degrees-of-freedom in terms of which we choose to resolve the problem.
The $\zeta$-evolution of hadron wave functions \cite{Lepage:1979zb, Efremov:1979qk} then ensures that parton splitting is properly expressed \cite{Ding:2019qlr, Ding:2019lwe}, undressing these compound functions and steadily exposing more of their partonic contents.

In terms of the two-particle LFWF in Eq.\,\eqref{LFWFexplicit}, the meson's valence $u$-quark DF is \cite{Brodsky:1989pv}
\begin{equation}
\label{PDFLFWF}
{\mathpzc u}^{M}(x;\zeta_H) = \int d^2k_\perp \, |\psi_{M_u}^{\uparrow\downarrow}(x,k_\perp^2;\zeta_H) |^2.
\end{equation}
$\bar{\mathpzc h}^{M}(x;\zeta_H) = {\mathpzc u}^{M}(1-x;\zeta_H)$ because $M$ is constituted solely from dressed $u$ and $\bar h$ degrees-of-freedom at $\zeta_H$.
A factorised expression of $\psi_{M_u}^{\uparrow\downarrow}(x,k_\perp^2;\zeta_H)$ is reliable for integrated quantities \cite{Xu:2018eii}, so it is sound to write
\begin{equation}
\psi_{M_u}^{\uparrow\downarrow}(x,k_\perp^2;\zeta_H) = \varphi_{M}^{u}(x;\zeta_H) \psi_{M_u}^{\uparrow\downarrow}(k_\perp^2;\zeta_H)\,,
\end{equation}
where $\psi_{M_u}^{\uparrow\downarrow}(k_\perp^2;\zeta_H)$ is a sensibly chosen function, \emph{viz}.\ its form is developed with careful attention to the physical nature of both the system under consideration and the intended applications.  Thus,
\begin{equation}
\label{PDFeqPDA2}
{\mathpzc u}^{M}(x;\zeta_H) \propto |\varphi_{M}^{u}(x;\zeta_H)|^2,
\end{equation}
with the proportionality constant fixed by the normalisation condition on $\psi_{M_u}^{\uparrow\downarrow}(x,k_\perp^2;\zeta_H)$.  Owing to parton splitting, \emph{i.e}.\ the evolution with energy scale of the active degrees-of-freedom, Eq.\,\eqref{PDFeqPDA2} is not valid on $\zeta>\zeta_H$.  Nevertheless, since the evolution equations for both DFs and DAs are known \cite{Dokshitzer:1977sg, Gribov:1972ri, Lipatov:1974qm, Altarelli:1977, Lepage:1979zb, Efremov:1979qk}, the connection changes in a traceable manner.

\smallskip

\noindent\textbf{3.$\;$Hadronic scale}.\,---\,%
One of the SM's key emergent phenomena is the appearance of a nonzero and finite gluon mass-scale as a consequence of highly-nonlinear, nonperturbative gauge sector dynamics \cite{Aguilar:2015bud, Gao:2017uox}.  The generation of this mass-scale does not alter any Slavnov-Taylor identities \cite{Taylor:1971ff, Slavnov:1972fg}; hence, all aspects and consequences of QCD's BRST invariance \cite{Becchi:1975nq, Tyutin:1975qk} are preserved.  The presence of this scale ensures that QCD's process-independent (PI) effective charge, $\hat{\alpha}(k^2)$, saturates in the infrared \cite{Binosi:2016nme, Cui:2019dwv}: $\hat{\alpha}(0)/\pi = 0.97(4)$.  An interpolation of the calculated result is provided by
\begin{align}
\label{Eqhatalpha}
\hat{\alpha}(k^2) & = \frac{\gamma_m \pi}{\ln\left[\frac{{\mathpzc K}^2(k^2)}{\Lambda_{\rm QCD}^2}\right]}
\,,\; {\mathpzc K}^2(y) = \frac{a_0^2 + a_1 y + y^2}{b_0 + y}\,,
\end{align}
%
$\gamma_m=4/[11 - (2/3)n_f]$, $n_f=4$, $\Lambda_{\rm QCD}=0.234\,$GeV, with (in GeV$^2$): $a_0=0.104(1)$, $a_1=0.0975$, $b_0=0.121(1)$.  
Here, $\Lambda_{\rm QCD}$ is the empirically determined mass-scale that arises in perturbation theory through the process of regularisation and renormalisation of QCD.

QCD's perturbative running coupling exhibits a (Landau) pole at $k^2=\Lambda_{\rm QCD}^2$.  This is unphysical \cite{Deur:2016tte}; and it is eliminated from the PI coupling by gauge sector dynamics: in Eq.\,\eqref{Eqhatalpha}, $k^2/\Lambda_{\rm QCD}^2 \to {\mathpzc K}^2(k^2)/\Lambda_{\rm QCD}^2$ as the logarithm's argument.  The value
\begin{equation}
m_G := {\mathpzc K}(k^2=\Lambda_{\rm QCD}^2) = 0.331(2)\,{\rm GeV}
\end{equation}
defines a screening mass.  It marks a boundary: the running coupling alters character at $k \simeq m_G$ so that modes with $k^2 \lesssim m_G^2$ are screened from interactions and the theory enters a practically conformal domain.  The line $k=m_G$ draws a natural border between soft and hard physics; hence, we identify $\zeta_H=m_G$.
Additional insights needed to connect perturbative and nonperturbative QCD are discussed, \emph{e.g}.\ in Refs.\,\cite{Binosi:2014aea, Deur:2016cxb}.

The hadronic scale, $\zeta_H$, is not directly accessible in analyses of experiments capable of providing information about DFs because certain kinematic conditions need to be met in order for the data to be interpreted in such terms \cite{Ellis:1991qj}.  These conditions require experiments with momentum transfers squared $Q^2 \sim \zeta_E^2 > m_N^2$.  Hence, any prediction for a DF at $\zeta_H$ must be evolved to $\zeta_E$ for comparison with experiment.  Following Refs.\,\cite{Ding:2019qlr, Ding:2019lwe, Cui:2019dwv}, we implement evolution by employing the PI effective charge in Eq.\,\eqref{Eqhatalpha} to integrate the one-loop DGLAP equations \cite{Dokshitzer:1977sg, Gribov:1972ri, Lipatov:1974qm, Altarelli:1977}.  This leads to an all-orders evolution scheme that enables predictions to be made for all $\pi$ and $K$ DFs. 

\begin{figure}[t]
\centerline{\includegraphics[clip, width=0.47\textwidth]{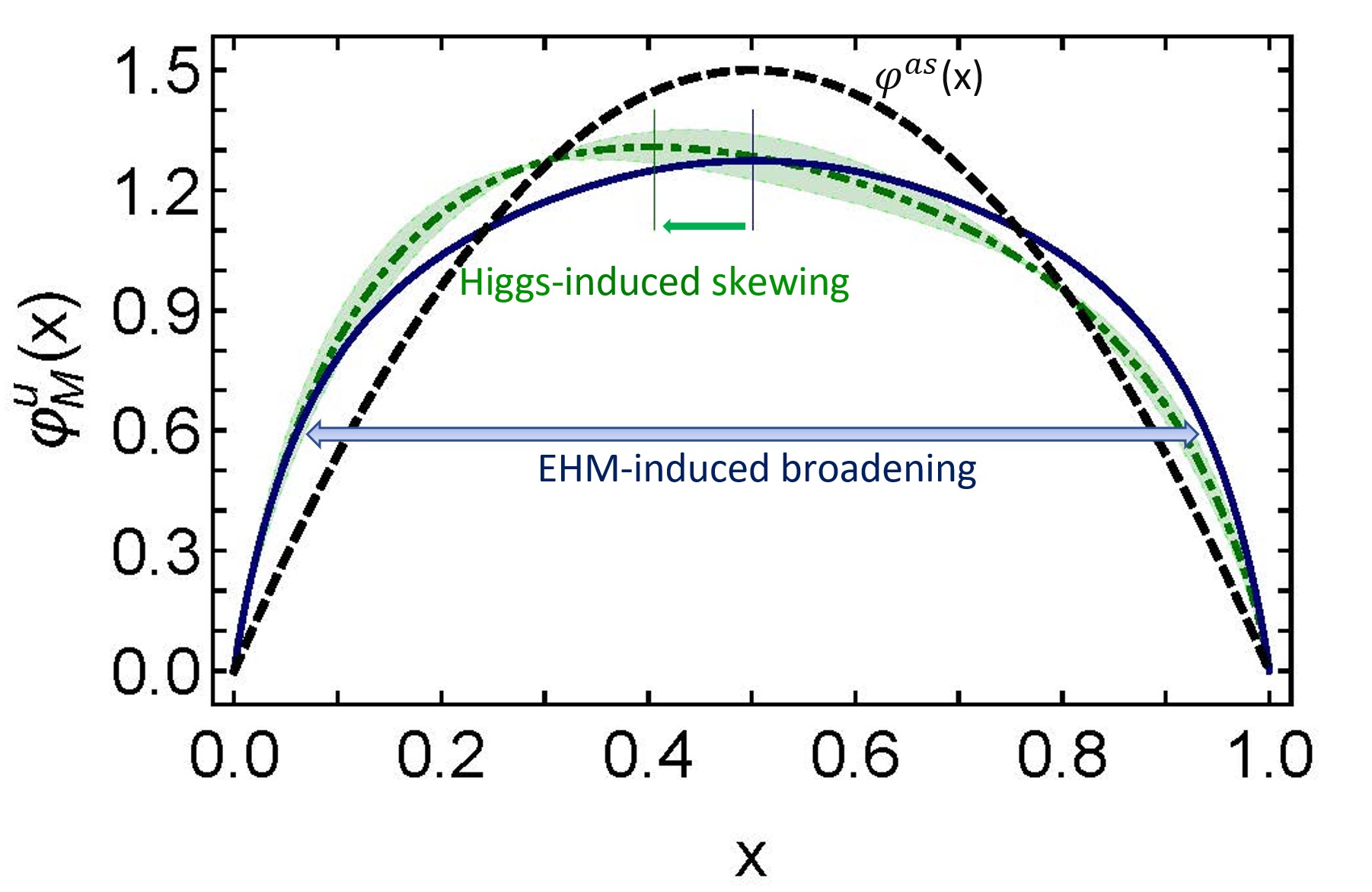}}
\caption{\label{FigPDAs}
DAs: pion, Eq.\,\eqref{pionDADB} -- solid blue curve; kaon, Eq.\,\eqref{NewKPDAForm} -- dot-dashed green curve within like-coloured bands; asymptotic DA, $\varphi^{\rm as}(x)=6 x (1-x)$ -- dashed black curve.  All ground-state meson DAs approach $\varphi^{\rm as}(x)$ as $m_N/\zeta \to 0$, where $\zeta$ is the energy scale associated with the given experiment.  However, at scales accessible in contemporary experiments, meson DAs are broadened as a consequence of EHM; and in systems defined by valence-quarks with different Higgs-produced current-mases, the peak is shifted away from $x=0.5$.}
\end{figure}

\smallskip

\noindent\textbf{4.$\;$Hadron scale distributions}.\,---\,%
Today, the pion's leading twist DA is now well constrained.  In comparison with the asymptotic profile \cite{Lepage:1979zb, Efremov:1979qk}: $\varphi^{\rm as}(x)=6 x (1-x)$, at all scales accessible in contemporary experiments, $\varphi_\pi^u$ is a broadened, flattened function \cite{Brodsky:2006uqa, Hwang:2012xf, Chang:2013pq, Zhang:2017bzy} -- see Fig.\,\ref{FigPDAs}.
A pointwise approximation to this DA is
\begin{align}
& \varphi_\pi (x;\zeta_H) = 20.227\, x(1-x) \nonumber \\
& \quad \times [1- 2.5088\, \sqrt{x(1-x)}+ 2.0250 \, x(1-x)]\,. \label{pionDADB}
\end{align}
(Eq.\,\eqref{pionDADB} updates the result in Ref.\,\cite{Chang:2013pq}, ensuring the DA's $x\simeq 0,1$ behaviour matches QCD's prediction.)

\begin{figure}[t]
\centerline{\includegraphics[clip, width=0.43\textwidth]{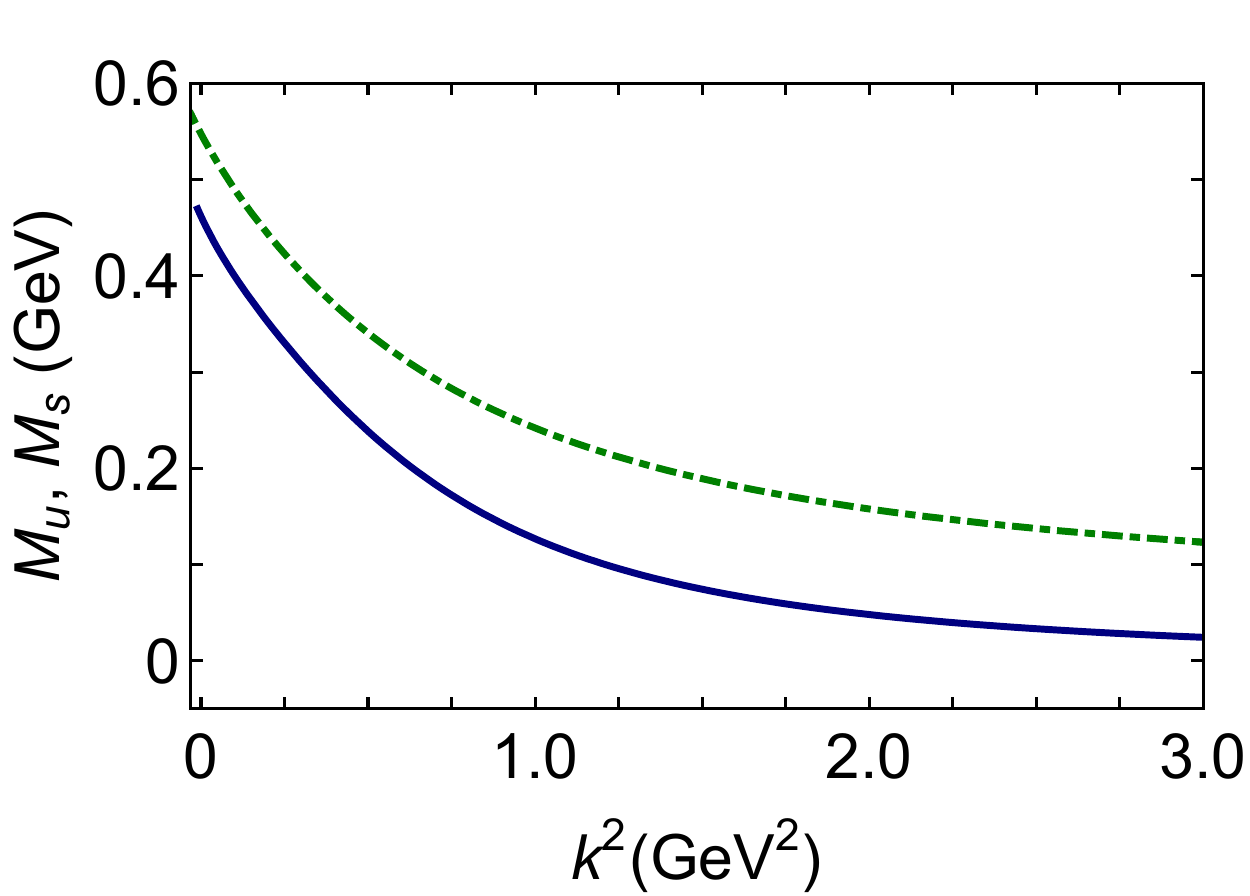}}
\caption{\label{FigMassFunctions}
Quark mass functions used to compute the pion and kaon DAs in Refs.\,\cite{Chang:2013pq, Shi:2014uwa}: $u=d$ quark -- solid blue curve; and $s$ quark -- dot-dashed green curve.
Whilst the ultraviolet behaviour is determined by the Higgs-generated current-quark masses, the infrared enhancement evident in both cases is EHM-driven: $M_u(0)=0.46\,$GeV, $M_s(0)=0.55\,$GeV.}
\end{figure}

The dilation of this DA is an important consequence of EHM, as expressed in the $u$-quark mass function depicted in Fig.\,\ref{FigMassFunctions}.  $M_u(0)=0.46\,$GeV is obtained by solving the gap equation using a Higgs-generated renorma\-lisation-group-invariant $u=d$-quark current-mass of just $\hat m_{u}=0.0043\,$GeV: \emph{i.e}.\ approximately 100-times smaller, $\hat m_u/M_u(0) \approx 0.01$.

Reflecting the lack of information about matter with strangeness, the kaon's DA is more uncertain \cite{Hwang:2012xf, Segovia:2013eca, Shi:2014uwa, Horn:2016rip, Gao:2017mmp}.
One can at most say that $\varphi_K(x;\zeta_H)$ is somewhat less broadened than $\varphi_\pi(x;\zeta_H)$ and also slightly asymmetric about $x=1/2$, both owing to the smaller role played by the Higgs-boson in connection with $u$ quarks as contrasted with $s$ quarks.

This is illustrated by the comparison between $u$- and $s$-quark mass functions in Fig.\,\ref{FigMassFunctions}: contrasting with the $u$-quark, $\hat m_s = 0.090\,$GeV, $M_s(0)=0.55\,$GeV, $\hat m_s/M_s(0) = 0.16$.
Moreover, $M_s(0)/M_u(0) =1.20$, in near agreement with the empirical value of the ratio of kaon and pion leptonic decay constants \cite{Zyla:2020}: $f_K/f_\pi=1.19$.  Inspection of Eq.\,\eqref{varphiresult} reveals that this link is not accidental, \emph{i.e}.\ (pseudo-)Nambu-Goldstone mode decay constants are an observable manifestation of the effective strength of mass generation summed over all sources \cite{Holl:2004fr, Krassnigg:2004if, Bhagwat:2006xi}.

Defining
\begin{equation}
\langle \xi^n = (1-2x)^n \rangle^{u_\zeta}_{M} = \int_0^1 dx\, (1-2x)^n\,\varphi_M^u(x;\zeta) \,,
\end{equation}
the preceding remarks can be stated as follows:
\begin{equation}
\langle [\xi,\xi^2]\rangle_\pi^{u_{\zeta_H}} = [0,0.25]\,,
\langle [\xi,\xi^2]\rangle_K^{u_{\zeta_H}}  = [0.035(5) , 0.24(1)]\,.
\label{DAmoments}
\end{equation}
Following established procedures \cite{Segovia:2013eca, Shi:2014uwa}, the moments in Eq.\,\eqref{DAmoments} can be used to determine the kaon's DA:
\begin{align}
\varphi_K^u&(x;\zeta_H) = {\mathpzc n}_{\varphi_K} \, x(1-x) \nonumber \\
& \times  \left[1 + \rho x^{\frac{\alpha}{2}} (1-x)^\frac{\beta}{2} + \gamma x^\alpha(1-x)^\beta\right]\,, \label{NewKPDAForm}
\end{align}
where ${\mathpzc n}_{\varphi_K}$ ensures unit normalisation, and the interpolation coefficients are listed in Table~\ref{parameterskaonDA}A.  Here ``upper'' indicates the curve that produces the largest value of $\langle\xi^2\rangle_K^{u_{\zeta_H}}$ and lower, the smallest.
Recall, $\varphi_K^{\bar s}(x;\zeta_H)= \varphi_K^{u}(1-x;\zeta_H)$.  The $K$ DA is drawn in Fig.\,\ref{FigPDAs}, where the shaded band expresses the uncertainty described above.

\begin{table}[t]
\caption{\label{parameterskaonDA}
(\textbf{A}) Coefficients and powers that specify the kaon DA determined by Eq.\,\eqref{NewKPDAForm}.  Upper, middle, lower refer to the values of $\langle \xi^2\rangle_{K}^{u_{\zeta_H}}$ produced by the identified coefficients.  
(\textbf{B}) Low-order moments of the kaon's ${\mathpzc u}$ and $\bar {\mathpzc s}$ DFs at $\zeta=\zeta_5$.  Mass-independent/dependent evolution is denoted by the subscript $\not \!\!m$/$m$, respectively.
}
\begin{center}
\begin{tabular*}
{\hsize}
{
l@{\extracolsep{0ptplus1fil}}|
c@{\extracolsep{0ptplus1fil}}
c@{\extracolsep{0ptplus1fil}}
c@{\extracolsep{0ptplus1fil}}
c@{\extracolsep{0ptplus1fil}}
c@{\extracolsep{0ptplus1fil}}}\hline\hline
(\textbf{A}) & ${\mathpzc n}_{\varphi_K}\ $ & $\rho\ $ & $\gamma\ $ & $\alpha\ $ & $\beta\ $\\\hline
{\rm upper}\ & $16.2\ $ & $4.92\ $ & $-6.00\ $ & $0.0946\ $ & $0.0731\ $\\
{\rm middle}$\ $&$18.2\ $ & $5.00\ $ & $-5.97\ $ & $0.0638\ $ & $0.0481\ $\\
{\rm lower}\ & $20.2\ $ & $5.00\ $ & $-5.90\ $ & $0.0425\ $ & $0.0308\ $\\
\hline\hline
\end{tabular*}

\smallskip

\begin{tabular*}
{\hsize}
{
l@{\extracolsep{0ptplus1fil}}|
c@{\extracolsep{0ptplus1fil}}
c@{\extracolsep{0ptplus1fil}}
c@{\extracolsep{0ptplus1fil}}}\hline\hline
(\textbf{B}) & $\langle x{\mathpzc q}^K \rangle\ $ & $\langle x^2{\mathpzc q}^K \rangle\ $ & $\langle x^3{\mathpzc q}^K \rangle\ $\\\hline
$u$ & $0.19(2)\ $ & $0.067(09)\ $ & $0.030(05)\ $\\
${\bar s}_{\not m}\ $ & $0.22(2)\ $ & $0.081(11)\ $ & $0.038(07)\ $ \\
${\bar s}_{m}\ $ & $0.23(2)\ $ & $0.085(11)\ $ & $0.040(07)\ $ \\\hline
$u+{\bar s}_{m}\ $ & $0.42(3)\ $ & $0.152(20)\ $ & $0.070(12)\ $ \\
\hline\hline
\end{tabular*}
\end{center}
\end{table}

Here it should be highlighted that the $\pi$ and $K$ DAs are unit-normalised.  In each case, an overall multiplicative factor of $f_{\pi,K}$, respectively, has been factorised.  Hence, what remains is an essentially local expression of EHM and Higgs-related interference effects.

Pion and kaon valence-quark DFs are now determined by Eq.\,\eqref{PDFeqPDA2}.
Glue and sea-quark distributions are identically zero at $\zeta_H$, being generated by DGLAP evolution on $\zeta>\zeta_H$ (Sec.\,3 and Refs.\,\cite{Ding:2019qlr, Ding:2019lwe}).  This expresses the feature that at $\zeta_H$ all glue and sea-quark contributions are sublimated into the dressed-quark and dressed-antiquark quasiparticles that emerge as solutions of the quantum field theory equations of motion.  The phenomenological success of constituent-quark models \cite{GellMann:1964nj, Zweig:1981pdFBS} rests on this sort of transmogrification.

\smallskip

\begin{figure}[t]
\centerline{\includegraphics[clip, width=0.34\textwidth]{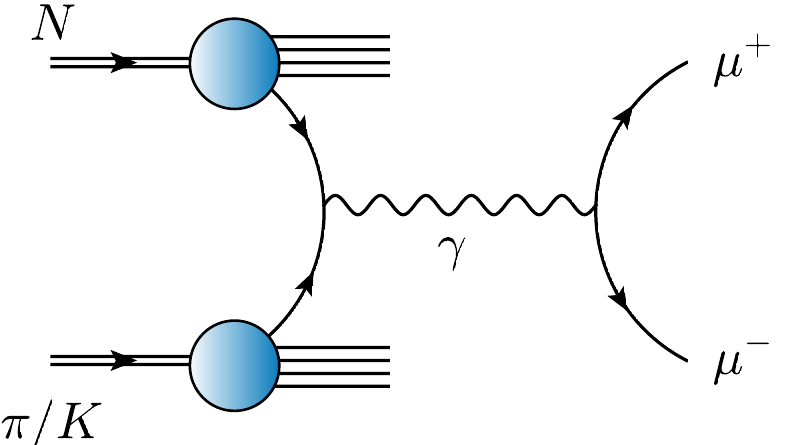}}
\caption{\label{FigDY}
Drell-Yan process: with appropriately chosen kinematics \cite{Drell:1970wh}, meson-nucleon collisions that produce lepton pairs with large invariant mass provide access to momentum distribution functions within the initial-state hadrons.}
\end{figure}

\noindent\textbf{5.$\;$Massless evolution}.\,---\,%
Regarding kaon structure functions, the only extant empirical information is the ratio ${\mathpzc u}^K(x)/{\mathpzc u}^\pi(x)$, extracted forty years ago \cite{Badier:1980jq} using the Drell-Yan process, depicted in Fig.\,\ref{FigDY}.  The mass-scale in this experiment was $\zeta \approx \zeta_5=5.2\,$GeV.  Thus, to deliver results for comparison with this data, our valence DFs must be evolved: $\zeta_H \to \zeta_5$.  Using the scheme explained in Sec.\,3, this yields DFs which produce
\begin{equation}
\langle x[{\mathpzc u}^K(x;\zeta_5) + \bar{\mathpzc s}^K(x;\zeta_5)] \rangle = 0.410\,,
\end{equation}
where $\langle x^n{\mathpzc q}^M\rangle = \int_0^1 dx x^n {\mathpzc q}^M(x;\zeta_5)$.  The same result is obtained by applying this procedure to ${\mathpzc u}^\pi(x;\zeta_H)$ defined via Eq.\,\eqref{PDFeqPDA2}.  Evidently, using mass-independent evolution, the kaon's valence-quark momentum fraction matches that in the $\pi$.

\begin{figure}[t]
\vspace*{2.7ex}

\leftline{\hspace*{2em}{\large{\textsf{A}}}}
\vspace*{-5ex}
\centerline{\includegraphics[clip, width=0.42\textwidth]{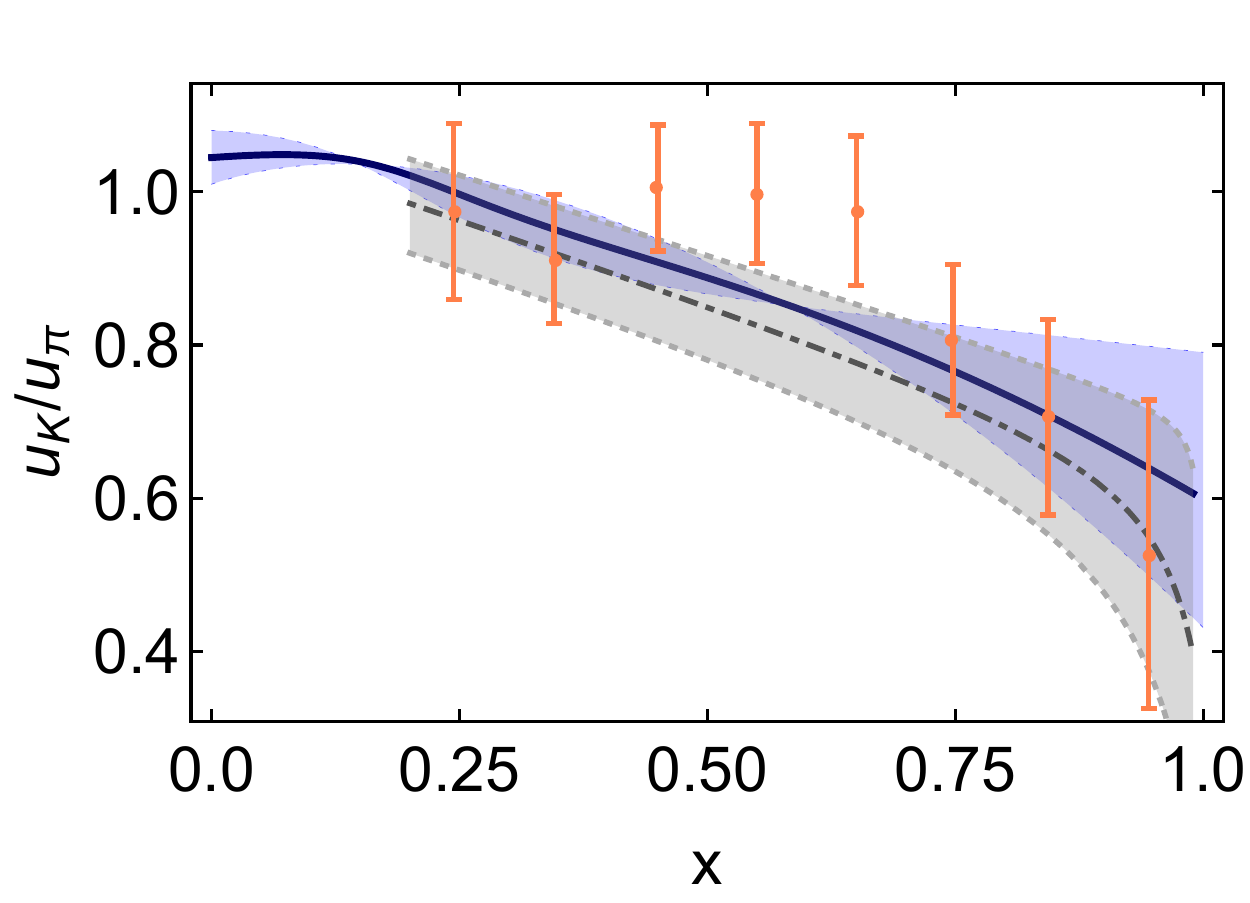}}
\vspace*{2ex}

\leftline{\hspace*{2em}{\large{\textsf{B}}}}
\vspace*{-5ex}
\centerline{\includegraphics[clip, width=0.42\textwidth]{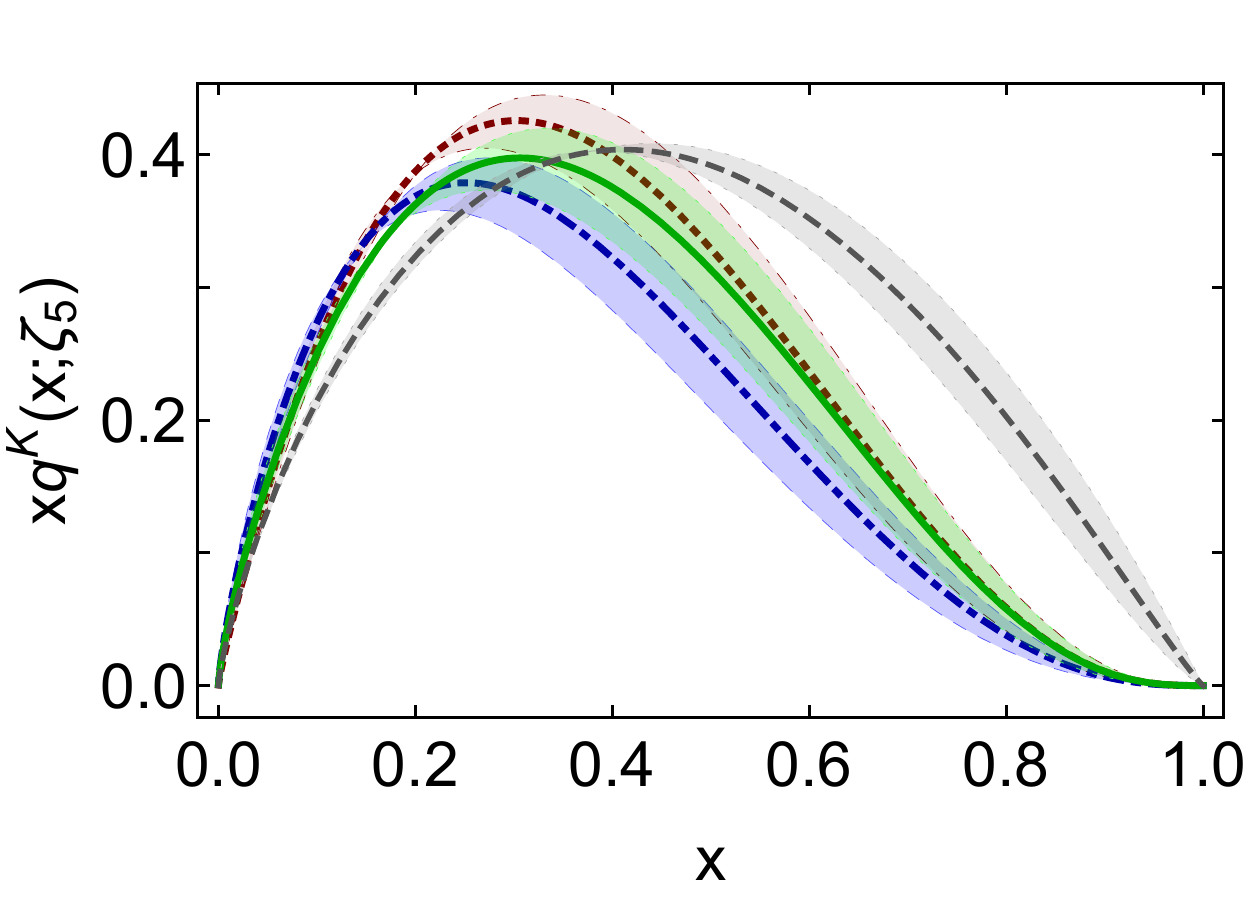}}
\caption{\label{qKzeta5}
(\textbf{A})
$u^K(x;\zeta_5)/u^\pi(x;\zeta_5)$ -- solid blue curve, our prediction ($\chi^2/$degree-of-freedom$\,=\,0.86$): the effect of $\zeta_H \to \zeta_H(1.0\pm 0.1)$ is negligible, producing no uncertainty larger than the linewidth.
The surrounding light-blue band marks the domain between the results obtained using Eqs.\,\eqref{PDFeqPDA2}, \eqref{NewKPDAForm}, Table~\ref{parameterskaonDA}A--upper and Table~\ref{parameterskaonDA}A--lower: ``lower'' produces the smallest $x=1$ value.
Dot-dashed grey curve within grey band -- lQCD \cite{Lin:2020ssv}: $\chi^2/$degree-of-freedom$\,=\,1.81$.
Data (orange) from Ref.\,\cite{Badier:1980jq}.
(\textbf{B})
$u^K(x;\zeta_5)$ - dot-dashed blue curve;
$\bar s_{\not m}^K(x;\zeta_5)$ [mass-independent splitting] -- solid green;
$\bar s_m^K(x;\zeta_5)$ [mass-dependent splitting] -- dotted maroon;
and dashed grey curve within grey bands -- lQCD result for $\bar s^K(x;\zeta_5)$ \cite{Lin:2020ssv}.  In this panel, the bands bracketing our central DF curves reflect the uncertainty in $\hat \alpha(0)$, Sec.\,3.
}
\end{figure}

Using the $\zeta_H \to \zeta_5$ evolved $\pi$ and $K$ DFs, one obtains the ratio $u^K(x;\zeta_5)/u^\pi(x;\zeta_5)$ drawn in Fig.\,\ref{qKzeta5}A.   The uncertainty existing in the kaon DA is expressed in the behaviour of this ratio on $x\gtrsim 0.5$: a broader kaon DA yields a ratio closer to unity at $x=1$.

The best agreement with data \cite{Badier:1980jq} is delivered by the kaon DF obtained using Eqs.\,\eqref{PDFeqPDA2}, \eqref{NewKPDAForm}, Table~\ref{parameterskaonDA}A--middle.  Hereafter, we focus on the kaon DFs defined by this curve; and  consider the impact of varying $\zeta_H\to (1.0\pm 0.1)\zeta_H$, thereby providing a conservative estimate of the uncertainty arising from that in the infrared value of the PI coupling, Sec.\,3.  This process yields the valence quark DFs plotted in Fig.\,\ref{qKzeta5}B, which produce the low-order moments in the first two rows of Table~\ref{parameterskaonDA}B.
Hence, accounting for $\zeta_H\to\zeta_H (1.0\pm 0.1)$,
$\langle x[{\mathpzc u}^K(x;\zeta_5) + \bar{\mathpzc s}^K(x;\zeta_5)] \rangle = 0.41(4)$.
Once again, this matches the pion result.


A first study of the kaon's valence-quark DFs using lattice-regularised QCD (lQCD) is now available \cite{Lin:2020ssv}.  It yields the following moments, listed according to the order in Table~\ref{parameterskaonDA}B:
$u$ \ldots\ $0.193(8)$, $0.080(7)$, $0.042(6)$; and
$\bar s$ \ldots\ $0.267(8)$, $0.123(7)$, $0.070(6)$.
These values are systematically larger than our predictions, especially for the $\bar s$.  The excesses are: $u$ \ldots\ $0.6(4.8)$\%, $21(6)$\%, $40(4)$\%; and $\bar s$ \ldots\ $24(7)$\%, $53(13)$\%, $84(16)$\%.  This is because, when compared with our predictions, the lQCD DFs are harder; a feature emphasised by Fig.\,\ref{qKzeta5}B.  Actually, the lQCD DF behaves as $(1-x)^{\beta}$, $\beta = 1.13(16)$, incompatible with the QCD prediction \cite{Ezawa:1974wm, Farrar:1975yb, Berger:1979du, Yuan:2003fs, Holt:2010vj, Chang:2013pq}:
\begin{equation}
\label{PDFQCD}
{\mathpzc q}^{M}(x;\zeta) \stackrel{x\simeq 1}{=} {\mathpzc c}_{\mathpzc q}^{M}(\zeta)\, (1-x)^{\beta(\zeta)}\,,\; \beta(\zeta)=2+\gamma(\zeta)\,,
\end{equation}
where ${\mathpzc c}_{\mathpzc q}^{M}(\zeta)$ is a constant and $\gamma(\zeta)$ increases logarithmically from zero on $\zeta>\zeta_H$.  Our predictions, on the other hand, are consistent with Eq.\,\eqref{PDFQCD}: $\beta(\zeta_5) = 2.73(7)$. 

Regarding the ratio ${\mathpzc u}^K(x;\zeta_5)/{\mathpzc u}^\pi(x;\zeta_5)$, the impact of $\zeta_H\to \zeta_H(1.0\pm 0.1)$ on both ${\mathpzc u}^\pi(x;\zeta_5)$ and ${\mathpzc u}^K(x;\zeta_5)$ is almost identical: it produces no uncertainty larger than the linewidth in Fig.\,\ref{qKzeta5}A.
The first lQCD results for this ratio are also drawn in Fig.\,\ref{qKzeta5}A.  The relative difference between the central lQCD result and our prediction is $\approx 5$\% despite the fact that the individual lQCD DFs are different from ours -- see Fig.\,\ref{qKzeta5}B.  This feature highlights that ${\mathpzc u}^K(x;\zeta_5)/{\mathpzc u}^\pi(x;\zeta_5)$ is forgiving of even large differences between the individual DFs used to produce the ratio.
More precise data is crucial before this ratio can be used effectively to test the modern understanding of SM NG modes; and results for ${\mathpzc u}^\pi(x;\zeta_5)$, ${\mathpzc u}^K(x;\zeta_5)$ separately have greater discriminating power \cite{Keppel:2015, C12-15-006A, Denisov:2018unjS}.

\smallskip

\noindent\textbf{6.$\;$Mass-dependent evolution}.\,---\,%
Hi\-therto, when implementing evolution, we used textbook forms for the massless splitting functions; so, the kaon's glue and sea distributions are practically the same as those in the pion.  Any symmetry-preserving study that begins at $\zeta_H$ with a bound-state built solely from dressed quasiparticles and enforces physical constraints on $\pi$, $K$ wave functions will produce this outcome when using massless splitting functions.  Naturally, the $\bar s$ quark is heavier than the $u$ quark.  Hence, \cite{Landau:1953um, Migdal:1956tc}: valence $\bar s$ quarks must generate less gluons than valence $u$ quarks; and gluon splitting must yield less $\bar s s$ pairs than light-quark pairs.  Such effects can be realised in the splitting functions.

We estimate the effect of mass-dependent evolution by modifying $s\to s$ and $g\to s$ splitting functions \cite{Chang:2014lva}:
{\allowdisplaybreaks
\begin{subequations}
\label{splittingfunctions}
\begin{align}
P_{s\leftarrow s}(z) & \to P_{q\leftarrow q}(z) - \Delta_{s\leftarrow s}(z,\zeta)\,, \label{Pss}\\
P_{s\leftarrow g}(z) & \to P_{s\leftarrow g}(z) + \Delta_{s\leftarrow g}(z,\zeta) \label{Pgs}\,,\\
\Delta_{s\leftarrow s}(z,\zeta) & = \sqrt{3}  (1 - 2 z) \sigma(\zeta)\,,\\
\Delta_{s\leftarrow g}(z,\zeta) & = \sqrt{5} (1-6 z+6 z^2)\sigma(\zeta)\,,
\end{align}
\end{subequations}}
\hspace*{-0.3\parindent}with $\sigma(\zeta) = \delta^2/[\delta^2 +(\zeta-\zeta_H)^2]$, $\delta = 0.1\,{\rm GeV} \approx M_s(0)-M_u(0)$, where $M_f(k^2)$ is the running mass of a $f$-quark.  (See Fig.\,\ref{FigMassFunctions}.)  All splitting constraints are preserved by Eqs.\,\eqref{splittingfunctions}.  The impacts of these modifications are clear: Eq.\,\eqref{Pss} reduces the number of gluons emitted by $\bar s$-quarks; and Eq.\,\eqref{Pgs} suppresses the density of $s\bar s$ pairs produced by gluons.  As required by the SM, both effects grow with quark mass difference, $\delta$, and decrease as $\delta^2/\zeta^2$ with increasing resolving scale.

\begin{figure}[t]
\vspace*{2.7ex}

\leftline{\hspace*{2em}{\large{\textsf{A}}}}
\vspace*{-5ex}
\centerline{\includegraphics[clip, width=0.42\textwidth]{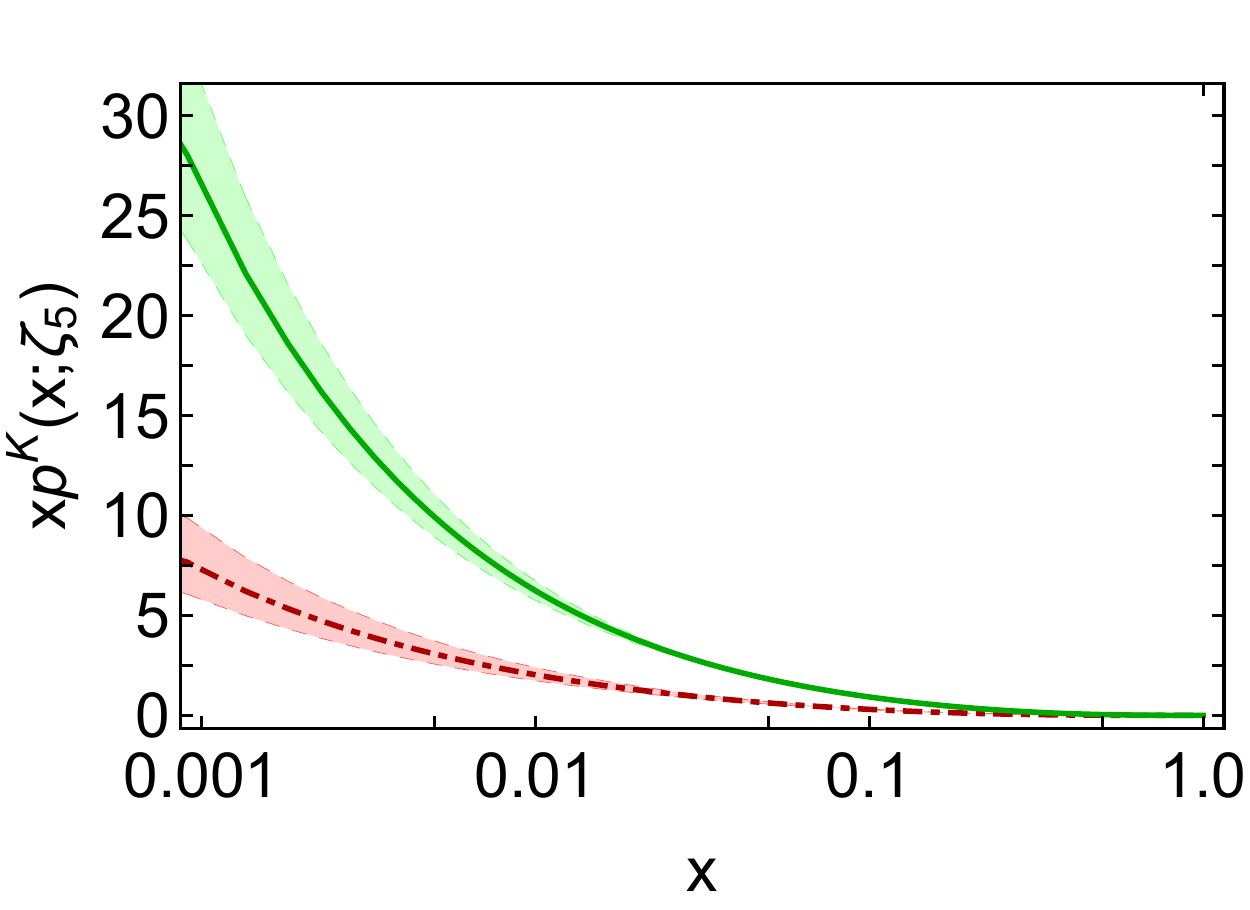}}
\vspace*{2ex}

\leftline{\hspace*{2em}{\large{\textsf{B}}}}
\vspace*{-5ex}
\centerline{\includegraphics[clip, width=0.42\textwidth]{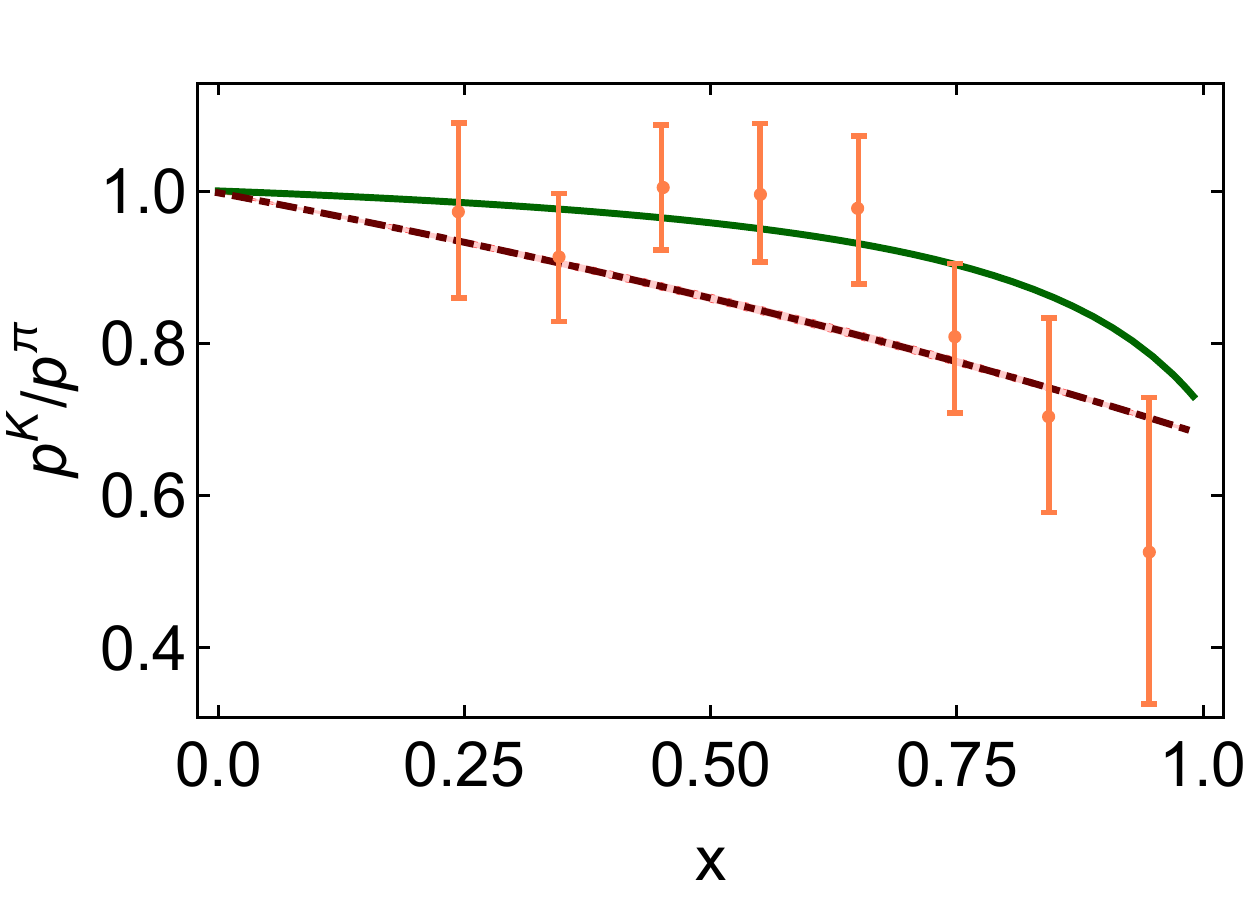}}
\caption{\label{kaongluesea}
(\textbf{A}) Solid green curve, $p=g$ -- our prediction for the kaon's glue distribution; and dot-dashed red curve, $p=S$ -- predicted kaon sea-quark distribution.
Normalisation convention: $\langle x[{\mathpzc u}^K(x;\zeta_5)+\bar{\mathpzc s}^K(x;\zeta_5)+{\mathpzc g}^K(x;\zeta_5)+S^K(x;\zeta_5)]\rangle=1$.
(\textbf{B}) Predictions: ${\mathpzc g}^K(x;\zeta_5)/{\mathpzc g}^\pi(x;\zeta_5)$ -- solid green curve;
and ${\mathpzc S}^K(x;\zeta_5)/{\mathpzc S}^\pi(x;\zeta_5)$ -- dot-dashed red curve.
Data on ${\mathpzc u}^K(x;\zeta_5)/{\mathpzc u}^\pi(x;\zeta_5)$ (orange) from Ref.\,\cite{Badier:1980jq} are included to guide comparisons.
(Shading in both panels displays the $\zeta_H \to \zeta_H(1.0\pm 0.1)$ uncertainty.)
}
\end{figure}

The new results for $\bar{\mathpzc s}^K(x;\zeta_5)$ are drawn in Fig.\,\ref{qKzeta5}B.  This DF produces the low-order moments in Rows~3 and 4 of Table~\ref{parameterskaonDA}B: the $u$-quark values are unchanged, but those for the $\bar s$-quark are increased by $4.8(8)$\%.

Our $\zeta=\zeta_5$ predictions for the kaon's glue and sea-quark DFs are drawn in Fig.\,\ref{kaongluesea}A.  These distributions vanish at $\zeta_H$ and are generated using the mass-dependent singlet evolution equations obtained following the procedure described in Sec.\,3 with the splitting function modifications in Eqs.\,\eqref{splittingfunctions}.  It is worth expressing these results via a comparison with pion glue and sea DFs.  Such ratios are depicted in Fig.\,\ref{kaongluesea}B.
%
Here the uncertainty owing to that in $\hat\alpha(0)$ is negligible, \emph{i.e}.\ no larger than the line width in either case.

The kaon's glue and sea-quark momentum distributions only differ from those of the pion on the valence region $x\gtrsim0.2$.  This is unsurprising because
mass-dependent splitting functions act principally to modify the valence DF of the heavier quark;
valence DFs are small at low-$x$, where glue and sea-quark distributions are large, and vice versa;
hence the chief impact of a change in the valence DFs must lie at large-$x$.
Notably, both of the predicted ratios in Fig.\,\ref{kaongluesea}B are similar to the measured value of ${\mathpzc u}^K(x;\zeta_5)/{\mathpzc u}^\pi(x;\zeta_5)$.
Conversely, the $K$ and $\pi$ glue and sea-quark DFs are practically identical on $x\lesssim 0.2$.
Using our DFs, we obtain ($\zeta=\zeta_5$):
$\langle x\rangle^K_g = 0.44(2)$,
$\langle x\rangle^K_{\rm sea} = 0.14(2)$,
with $\langle x\rangle^K_{{\rm sea}_{{\mathpzc l}}} = 0.091(11)$, $\langle x\rangle^K_{{\rm sea}_{s}} = 0.045(06)$, where ${\mathpzc l}$ denotes the light-quarks ($u+d$).
Comparing these results with those for the pion, then accounting for mass-dependent splitting functions, we find that the gluon light-front momentum fraction in the kaon is $\sim 1$\% less than that in the pion and the sea fraction is $\sim 2$\% less.

\smallskip

\noindent\textbf{7.$\;$Outlook}.\,---\,%
The standard model of particle physics (SM) has two mass-generating mechanisms.  That connected with the Higgs boson is understood within perturbation theory.  On the other hand, so far as everyday matter is concerned, the Higgs mechanism is directly responsible for $\lesssim 2$\% of measurable mass.  Present-day theory indicates that the remaining $\gtrsim 98$\% emerges from strong interactions within the SM.  Elucidation of the manifold consequences of this emergent mass and the nature of interference effects between these two mass producing mechanisms are the focus of worldwide attention in high-energy nuclear and particle physics because they are basic to understanding the evolution of the Universe.  In approaching these questions, far-reaching insights are expected to be revealed by charts of the structure of $\pi$- and $K$-mesons and their subsequent comparison with similar maps for baryons.

Herein, we delivered predictions for all $\pi$, $K$ distribution functions: valence, glue and sea-quark.  Regarding $\pi$ and $K$ valence-quark distributions, new algorithms for the application of lattice-regularised QCD are delivering preliminary results; but many challenges remain before quantitative precision can be achieved therewith.  Hence, our predictions for the entire array stand alone: they reveal the dominant role of emergent hadronic mass in forming all DFs and the impact of Higgs-boson modulations in distinguishing between $\pi$ and $K$ structure.

Our analysis can be improved in two ways: one could further test the factorisation assumption made for meson light-front wave functions; and a more rigorous treatment of mass-dependence in splitting functions should be implemented.  Both improvements are underway.

The SM's (pseudo-) Nambu-Goldstone modes are basic to the formation of everything, from nucleons to nuclei, and on to neutron stars.  Hence, new-era experiments capable of testing the predictions herein should have high priority \cite{Keppel:2015, C12-15-006A, Denisov:2018unjS, Chen:2020ijn}.  Equally, the phenomenological methods used to proceed from data to DFs must match experiments in precision.  In that connection, it is anticipated that this study will also stimulate progress.

\smallskip

%
\noindent\textbf{Acknowledgments}.\,---\,%
We are grateful for constructive comments from
V.~Andrieux, P.~Barry, W.-C.~Chang, C.~Chen, O.~Denisov, J.~Friedrich, W.~Melnitchouk, C.~Mezrag, W.-D.~Novak, S.~Platchkov, M.~Quaresma and C.~Quintans; and
for the hospitality and support of RWTH Aachen University, III.\,Physikalisches Institut B, Aachen - Germany.
Work supported by:
National Natural Science Foundation of China (grant no.\,11805097),
Jiangsu Provincial Natural Science Foundation of China (grant no.\,BK20180323).
Jiangsu Pro\-vince \emph{Hundred Talents Plan for Professionals};
Alexander von Humboldt Foundation;
and
Spanish Ministry of Science and Innovation (MICINN), grant no.\ PID2019-107844GB-C22.
%




\end{document}